\documentclass[twocolumn,secnumarabic,amssymb, nobibnotes, aps, prd]{revtex4-1}

\usepackage{graphicx}
\usepackage{hyperref}
\usepackage{amsmath}
\usepackage{xcolor}

\begin{document}
\color{red}
\title{Einstein's cluster mimicking compact star in the teleparallel equivalent of general relativity }%

\author{Ksh. Newton Singh}%
\email[Email:]{ntnphy@gmail.com}
\affiliation{Department of Physics, National Defence Academy, Khadakwasla, Pune-411023, India. \\
and\\
Faculty Council of Science, Jadavpur University, Kolkata-700032, India}

\author{Farook Rahaman}%
\email[Email:]{rahaman@associates.iucaa.in}
\affiliation{Department of Mathematics, Jadavpur University, Kolkata-700032, India}

\author{Ayan Banerjee }%
\email[Email:]{ayan\_7575@yahoo.co.in}
\affiliation{Department of Mathematics, Jadavpur University, Kolkata-700032, India}

\date{today}%

\begin{abstract}

We present a physically plausible solution representing Einstein's cluster mimicking the behaviors of compact star in the context of teleparallel equivalent of general relativity. The Teleparallel gravity (TEGR) is an alternative formulation of gravity which uses tetrads as the dynamical variables. We focus on two particularly interesting scenarios. First, we develop the Einstein clusters in TEGR field equations using effective energy-momentum tensor for diagonal as well as off-diagonal tetrad. We then study the clusters in modified $f(T)-$gravity for anisotropic fluid distribution. Based on these two theories, we further study  the solution without net electric charge and then for charged solution. For charge parameter $k \rightarrow 0$, the charged solution reduces to neutral one. Our calculations show that when charge increases, the stiffness of the EoS also increases. This is due to increase in adiabatic index and sound speed approaching speed of light. When the charge increase beyond a certain limit ($0\le k \le 1.3 \times 10^{-5}$ and $0\le k \le 1 \times 10^{-6}$), the compactness parameter crosses the Buchdahl limit i.e. $2M/R>8/9$ and the solution start violating the causality condition. We test the Tolman-Oppenheimer-Volkoff (TOV) limit for such compact objects. We analyze the static stability criterion of the Einstein clusters for both charged and uncharged case, and the stability of such compact objects is enhanced by the presence some net electric charge. In addition, we present and discuss the energy conditions, causality condition and the adiabatic index close to the stability limit. After analyzing these problems, we conclude that the Einstein clusters do exists only if $f(T)$ is a linear function of the torsion scalar $T$, that is in the case of Teleparallel Equivalent of General Relativity.  Finally, we compare our solution for pure general relativity. As a result, we concluded that the Einstein cluster solution do exist in pure GR, however, physically unfit to mimic compact stars. We have also extend our findings by assuming the diagonal or off-diagonal tetrad and specific case of $f(T)$. In such models, Einstein's cluster solutions do exist however can't mimic the properties of a compact star.

\end{abstract}


\maketitle

\section{Introduction}

The concepts of Einstein's cluster \cite{ein} was introduced in 1939 to understand the system of stationary gravitating particles each moving along circular path about a common center under the influence of their combine gravitational field. If these particles are orbiting in same path with different phases or a similar orbit but inclined at a different angle, it can construct a shell named as ``Einstein's Shell". By constructing layers of Einstein's shell a \textit{Einstein's Cluster} is formed. All these particles are distributed spherically symmetric in sufficient continuous, random and collisionless geodesics. Such systems are static and in equilibrium where the gravitational force is balanced by the centrifugal force. Therefore, in this way a thick spherical shell of matter is composed without pressure in the radial direction, but only tangential stresses.

Einstein clusters have been extensively studied in different literature \cite{flo,zap,gil,Comer1}. Specially, the  components of energy momentum tensor for the cluster have been obtained in several representations. Due to the spherical
symmetry nature, the only non-vanishing components of $T^{\mu}_{\nu}$ are $T^{t}_{t}= \rho^{\text{eff}}$,  $T^{r}_{r}= -p_r^{\text{eff}}$ and $T^{\theta}_{\theta}= T^{\varphi }_{\varphi }= -p_t^{\text{eff}}$, with $\rho^{\text{eff}}$ is the effective energy density and  $p_r^{\text{eff}}$ and $p_t^{\text{eff}}$ are the effective radial and tangential pressures of the cluster, respectively. In principle, the junction conditions
require $p_r^{\text{eff}}$ to be continuous across the boundary of each layer of the shell, which follows that for the Einstein cluster $p_r^{\text{eff}}=0$. Therefore, Einstein's clusters are known for their highly anisotropic in nature, for which the radial pressure is different from the tangential one, $p_r^{\text{eff}} \neq p_t^{\text{eff}}$.

Interesting features of relativistic anisotropic matter distributions have been extensively pointed out as early as 1933 by Lema\^{i}tre
\cite{Lema,Lema1}. Though in ref.\cite{Bowers} Bowers and Liang build up anisotropic models as the beginning of the epoch of more active research. In the latter, anisotropic model has widely been studied in many astrophysical objects such as stars, gravastars, wormholes etc. Not surprisingly, in the last few decades there has been renewed interest in structure and evolution of compact objects with interior anisotropic fluids (see for instance \cite{Mak:2001eb,Herrera:2004xc,Chaisi:2005rb,Abreu:2007ew,Thirukkanesh:2008xc,Maurya:2015maa,Folomeev:2015aua,Kalam:2012sh,Maurya:2018kxg,Bhar:2017hbw,NewtonSingh:2017nls,Bhar:2016xnl}). In \cite{Boehmer:2006ye,Andreasson:2009pe}, upper and lower bounds for spherically-symmetric static solutions of the Einstein-fluid equations in presence of a positive cosmological constant.

 Within the context of GR, Florides \cite{flo} had tried to understand why a spherically symmetric distribution of pressure-less dust at rest cannot maintain itself in equilibrium. Since this attempt had opened up a new interior uniform density (Schwarzschild-like) solution. Specifically, the obtained interior solution process a positive tangential pressure which is increasing  function of the radial coordinate and having constant density. In next, it has been found that Florides interior solution describes the interior field of an \textit{Einstein cluster}. In this spirit, a modified approach to the problem of relativistic clusters was proposed by Zel'dovich and Polnarev \cite{Zel}. Further,  Zapolsky \cite{zap} discussed the stability of such clusters by adopting the same methods which is used to study the stability of compact stars. Gilbert \cite{gil} investigated the stability of Einstein clusters, leading to an upper limit to the velocities of the particles in the cluster. In this same context,  Cocco and Ruffini \cite{Cocco} introduced the concept of metastable clusters by considering explicit examples.

In the 1970's, Hogan \cite{Hogan} who suggested that neutrinos can be emitted from Einstein's cluster at a very specific angle ranges from 0 to $\pi/2$ with the radial direction. If the cluster is made up of charged particles than  it can't be in stationary equilibrium so long as the total charge is greater than or equal to the total mass \cite{Banerjee}. Interestingly, Einstein's cluster is also considered as a spin fluid with zero pressure where the spin density vanishes at the boundary  of the fluid sphere \cite{Bedran}. In a sense, elliptical Einstein  shells by means of ``elliptical" orbits was studied \cite{Comer}, however,  such configurations were not stable and eventually reduces  to spherical shells (see \cite{Comer} for review).

Number of modified gravity theories have been proposed which may describe the accelerated expansion of the universe in an effective level. This endeavor arises from unifying gravitation and quantum mechanics, and addressing some cosmological problems which include the dark energy problem (non-standard cosmic fluid with negative pressure) in the late Universe and singularity problem in the early Universe. In addition to theoretical considerations  modified gravity also could give adequate description of cosmological observations \cite{Bahcall:1999xn,Bamba:2012cp,Joyce:2014kja,Perrotta:1999am}. A systematic review on recent progress in the construction of modified gravity models has been done
(see Nojiri \emph{et al} \cite{Nojiri:2017ncd}) in cosmology, emphasizing on inflation, bouncing cosmology and late-time acceleration era.

Among the modified theories of gravity, recently $f(T)$ gravity has attracted much attention in the community. Inspired by the formulation of $f(R)$-gravity \cite{Sotiriou:2008rp,Santos:2008qp,Harko:2008qz,Capozziello:2018wul,Astashenok:2014gda,Astashenok:2013vza,Goswami:2014lxa}, where $f(R)$ is a generic function  of the Ricci scalar $R$ of the underlying geometry, $f(T)$ gravity is a similar generalization. This theory is based on the old definition of the ``Teleparallel equivalent to General Relativity" (TEGR) \cite{Hayashi,Hayashi1,Hayashi2}, where the Lagrangian is an analytic function of the torsion scalar $T$ \cite{Ferraro:2008ey,Fiorini:2009ux}. The basic equations of GR and its teleparallel equivalent is $R = - T +B$, where $R$ and $T$ are the Ricci scalar and torsion scalar with $B=\frac{2}{e}\partial_{\mu}(eT^{\mu})$ is a total derivative term which only depends on torsion.
Thus, Einstein-Hilbert action can now be represented in two distinct ways, either using the Ricci scalar or the torsion scalar, and consequently these two theories have the same equations of motion \cite{Bahamonde:2015hza,Bahamonde:2015zma}. However, the theoretical framework of $f(T)$ gravity depends on an appropriate ansatz for the tetrad field. It is interesting to mention that this theory is not invariant under local Lorentz transformations, and therefore the choice of tetrad plays an important role in determining such model. However TEGR, as a torsion theory, is equivalent to GR, but $f(T)$ theory is not equivalent to GR.

Although, $f(T)$ gravity does not coincide with $f(R)$ gravity.  The main catch point is that for a nonlinear $f(R)$ function, gravity is a fourth-order theory, whereas $f(T)$-gravity field equations are always second-order. {
At this point one should have noted that $f(R)$ modified gravity, in the Palatini version,
can also be viewed as a second order system of equations \cite{Durrer:2008in,DeFelice:2010aj}. Compare to $f(R)$ gravity, the action of $f(T)$ theory and the field equations are not invariant under local Lorentz transformations \cite{Li:2010cg}, which relates to the fact that
$f(T)$ theories appear to have extra degrees of freedom with respect to the teleparallel equivalent of GR \cite{Sotiriou:2010mv},
although their physical nature is not yet well understood.} Though, there has been a growing interest in this kind of theories due to its ability to explain both early \cite{Bamba:2016wjm,Jamil:2013nca,Qiu:2018nle}, as well as at late times accelerating phases of the Universe  \cite{Paliathanasis:2016vsw,Hohmann:2017jao,Cai:2015emx,Capozziello:2018hly} without the inclusion of a dark energy fluid.

In recent years attentions have been focused on the  gravitational waves from compact binary \cite{Nunes:2019bjq,Nunes:2018evm}  in $f(T)$ gravity.
 {In \cite{Bamba:2014zra},  the effects of the trace anomaly on inflation in $T^2$ gravity
has been examined.}
On the other hand, this model has been used for studying wormhole solution (see \cite{Yousaf:2017hjh,Rani:2016gnl,Rani:2016zbd} and references therein). The structure of compact stars in $f(T)$ gravity was investigated recently in refs. \cite{Ilijic:2018ulf}. This method is examined for $f(T)$ theory where a special form of $f(T)= T+ {\alpha \over 2} T^2$ is selected. Similar studies have also concluded that due to presence of anisotropic fluid affects the value of luminosities, redshifts, and maximum mass of a compact relativistic object in \cite{Abbas:2015yma,Momeni:2016oai,Abbas:2015zua}.

Recently, Lake \cite{lak} and Bohmer \& Harko \cite{Boehmer:2007az} considered Einstein's cluster of WIMPs dark matter generating spherically symmetric gravitational field of a galactic halo that can fit the rotational curve of any galaxy by adjusting two parameters (i) angular momentum distribution and (ii) number distribution of the WIMPs. Also it was  shown that Einstein's clusters were dynamically stable under radial and non-radial perturbations \cite{Boehmer:2007az}. The gravitation lensing due to such Einstein's cluster is slightly smaller as compare to isothermal sphere of dark matter \cite{Boehmer:2007az}.

Inspired by the above applications of teleparallel and $f(T)$ theories of gravity, we are interested to investigate solution representing Einstein's cluster. The manuscript is organized as follows: in Sec.~\ref{sec1} we briefly review the foundations of  teleparallel and $f(T)$ theories. We find the corresponding field equations for general spherically symmetric spacetime with diagonal and off-diagonal tetrad, and  by assuming different $f(T)$ function. In Sec.~\ref{sec2}, we derive   charged and uncharged solutions for Einstein cluster and compare them with standard GR model.
 In Sec.~\ref{sec2}, we derive the Einstein clusters for charged and uncharged solutions.  The metric exterior to the sphere is given by Reissner-Nordstr$\ddot{o}$m metric in Sec.~\ref{sec3}. Sec.~\ref{sec4} and Sec.~\ref{sec5}  are devoted to discuss the stability of the Einstein cluster model. The modified Oppenheimer-Volkoff limit is analyzed as well as other properties of the spheres, such as causality condition, adiabatic index. Moment of inertia and time period of the cluster are obtained in Sec.~\ref{sec6}. Finally, in Sec.~\ref{sec7} we summarized the results.

\section{Field equation and spherically symmetric solutions in $f(T)$ gravity}\label{sec1}

In this section we briefly present the main points of the $f(T)$ gravity. We get the field equations by varying the action
\begin{eqnarray}\label{eq1}
S=\int d^4x~\left[{f(T) \over 16\pi}+\mathcal{L} \right],
\end{eqnarray}
where $\mathcal{L}$ is the matter Lagrangian with $G=c=1$, and $T$ is the torsion scalar constructed from the torsion tensor:
\begin{eqnarray}\label{eq2}
T^\sigma_{\mu \nu} &=& \Gamma^\sigma_{\mu \nu}-\Gamma^\sigma_{\nu \mu} = e_i^\sigma \big(\delta_\mu e^i_\nu-\delta_\nu e^i_\mu \big).
\end{eqnarray}
Notably, the difference of Weitzenb$\ddot{\text{o}}$ck connection and the Levi-Civita connection $\widetilde{T}^\sigma_{\mu \nu}$ widely used in GR is defined as the
contorsion tensor $K^{\mu \nu}_\sigma$ as follows
\begin{eqnarray}\label{eq3}
K^{\mu \nu}_\sigma &\equiv & T^\sigma_{\mu \nu}-\widetilde{T}^\sigma_{\mu \nu} \\ \nonumber
&=& {1 \over 2} \Big(T^{\mu \nu}_\sigma + T^{\nu \mu}_\sigma - T^{\mu \nu}_\sigma \Big).
\end{eqnarray}
In $f(T)$ geometry, we introduce for convenience, the ``superpotential", namely
\begin{eqnarray}\label{eq4}
S^{\mu \nu}_\sigma &=& K^{\mu \nu}_\sigma-\delta^\nu_\sigma ~T^{\alpha \mu}_\alpha+\delta^\mu_\sigma~ T^{\alpha \nu}_\alpha,
\end{eqnarray}
and then the torsion scalar $T$ is given by
\begin{eqnarray}\label{eq5}
T \equiv T^\sigma_{\mu \nu}S^{\mu \nu}_\sigma,
\end{eqnarray}
which is equivalent to the Ricci scalar $R$ up to a total derivative term. Actually, with the existence of torsion tensor, in particular, if $f(T) = T$, the resulting equations of motion are equivalent to GR, and $T^\sigma_{\mu \nu}$ no longer be expressed in terms of metric, but act as an independent variable (see reviews \cite{Hehl,Cai}).

In torsional formulations of gravity ones uses the tetrad fields $e^i_\mu$ are related to the  metric tensor $g_{\mu \nu}$ by $g_{\mu \nu} (x)= \eta_{ij} e^i_\mu(x)e^j_\nu(x)$, where $\eta_{ij}$ is the Minkowski metric of the tangent space with the form of $\eta_{ij}= \text{diag}~(1,-1,-1,-1)$.

Additionally, we define the co-tetrad $e^i_\mu$ through
\begin{eqnarray}\label{eq6}
e^\mu_i e^i_\nu=\delta^\mu_\nu ~~ \text{and} ~~ e^\mu_i e^j_\mu=\delta^j_i,
\end{eqnarray}
with $e=\sqrt{-g}=det(e^i_\mu)$. Variation of the action (\ref{eq1}) with respect to the tetrad field gives the field equations
\begin{eqnarray}\label{eq7}
 S^{\mu \nu}_i f_{,TT}~T_{,\mu} + e^{-1} (eS^{\mu \nu}_i)_{,\mu}~f_{,T} \nonumber \\
 - T^\sigma_{\mu i} S^{\nu \mu}_\sigma f_{,T}
 {1\over 4} e^\nu_i f = 4\pi \mathcal{T}^\nu_i .
\end{eqnarray}
where $f_{,T}$ and $f_{,TT}$ denote the first and second derivatives of the function $f(T )$ with respect to $T$, and the tensor $\mathcal{T}^\nu_i$ represents the energy-momentum tensor of the matter source $\mathcal{L}$. Considering the description of energy momentum tensor, which, in the present study is written as $\mathcal{T}^\nu_i$ = $M^\nu_i +E^\nu_i$. Since, $M^\nu_i$ stands for the energy-momentum tensor of an anisotropic fluid distribution and $E^\nu_{i}$ is the electromagnetic energy-momentum tensor. So, the complete form of Einstein-Maxwell field equations is

\begin{eqnarray}
M^\nu_i &=& (p_t+\rho)u^\nu u_i - p_t g^\nu_i + (p_r-p_t)\chi_i \chi^\nu,\\ \label{eq8}
E^\nu_i  &=& {1 \over 4\pi} \left({1 \over 4\pi}~g^\nu_i F_{\alpha \beta} F^{\alpha \beta}- g^{\alpha \beta} F^\nu_\alpha F_{i \beta}\right),\label{eq9}
\end{eqnarray}
where $u_\nu$ is the four-velocity and $\chi_\nu$ is the unit spacelike vector in the radial direction. Moreover, the electromagnetic tensor $F_{ij}$ satisfies Mexwells equations
\begin{eqnarray}\label{eq10}
 F_{\alpha \gamma,\beta}+F_{ \gamma \beta, \alpha}+F_{ \alpha \beta,\gamma}= 0,\nonumber \\
 \left[\sqrt{-g}F^{\alpha \beta}\right]_{,\beta}=4\pi J^{\alpha}\sqrt{-g},
 \end{eqnarray}
where  $J^{\alpha}= \sigma u^{\alpha}$  is the electric current density and $F_{\alpha \beta}$ denotes the skew symmetric electromagnetic field tensor, with the parameter $\sigma$ is the charge density.

Since we are interested in spherically symmetric solutions that can be used to describe  a dense compact relativistic star. To this end, we write the space-time metric in  the following form
\begin{equation}\label{eq11}
ds^{2}=e^{\nu(r)}dt^{2}-e^{\lambda(r)}dr^{2}-r^{2}\left(d\theta^{2}+\sin^{2}\theta ~d\phi^{2} \right),
\end{equation}
where  $(t, r, \theta, \phi)$ are the usual Schwarzschild-like coordinates, with  $\nu$ and $\lambda$ are the functions of the radial coordinate $r$, are yet to be determined. Now, by considering diagonal and off-diagonal tetrad with different functional forms, we derive different field equations, as $f(T)$ theory
is not invariant under local Lorentz transformations.

\subsection{Diagonal tetrad and $f(T)=aT+b$}

Here, we start with the simplest possible diagonal tetrad (T1) giving this metric (\ref{eq1}) as follows \cite{Abbas:2015yma,Momeni:2016oai}:
\begin{equation}\label{eq12}
[e^i_\mu]= \text{diag}(e^{\nu/2},~e^{\lambda/2},~r,~r \sin \theta),
 \end{equation}
 and its determinant is $|e^i_\mu|=r^2 \sin \theta ~ e^{(\nu+\lambda)/2}$. The corresponding torsion scalar and its derivative  is given by
 \begin{eqnarray}
T(r) &=& {2e^{-\lambda} \over r}\left(\nu'+{1 \over r} \right),\\ \label{eq13}
T'(r) &=&  {2e^{-\lambda} \over r}\left[\nu''+{1 \over r^2}- T\left(\lambda'+{1 \over r} \right)\right], \label{eq14}
\end{eqnarray}
where the prime denotes the derivative with respect to $r$. Thus, the general field Eq. (\ref{eq7}) give rise to the explicit equations of motion:
\begin{eqnarray}
4\pi \rho+ E^2 &=& {f \over 4} - {f_{,T} \over 2}\left[T-{1 \over r^2}-{e^{-\lambda} (\lambda'+\nu') \over r} \right], \label{eq15}\\
4\pi p_r-E^2 &=& {f_{,T} \over 2}\left( T-{1 \over r^2}\right)-{f \over 4},  \label{eq16} \\
4\pi p_t+E^2 &=& \left[{T\over 2}+e^{-\lambda}  \left\{{\nu'' \over 2}+\left({\nu'\over 4}+{1 \over 2r} \right) \big(\nu'-\lambda'\big) \right\} \right]  \nonumber \\
&& \times {f_{,T} \over 2}-{f \over 4}, \label{eq17}\\
{\cot\theta \over 2r^2 }~ T'~f_{,TT} &=& 0, \label{eq18} \\
\sigma (r) &=& {e^{-\lambda/2} \over 4\pi r^2} ~\big(r^2 E\big)'.  \label{eq19}
\end{eqnarray}
The Eq. (\ref{eq18}) puts a strict constraint on the possible $f(T )$ functions. We immediately observe  few interesting facts. The Eq. (\ref{eq18}) implies here that all solutions satisfy either $f_{,TT} = 0$ or $T' = 0$, where the former reduces the theory to TEGR (see Ref. \cite{Boehmer:2011gw}). As a result, the choice of $f_{,TT} = 0$ leads to the following linear model \cite{Abbas:2015zua}:
\begin{equation}\label{eq20}
   f(T)= aT+b,
\end{equation}
where $a$ and $b$ are integration constants. Inserting Eq. (\ref{eq20}) into the field Eqs. (\ref{eq15}-\ref{eq17}), one can obtain the modified field equations in Teleparallel gravity as,
\begin{eqnarray}
4\pi \rho +E^2 &=& \frac{2 a \left(r e^{-\lambda} \lambda '-e^{-\lambda}+1\right)+b r^2}{4 r^2}, \label{eq21}\\
4\pi p_r -E^2 &=& \frac{2 a e^{-\lambda} \left(r \nu'+1\right)-2 a-b r^2}{4 r^2}, \label{eq22}\\
4\pi p_t +E^2 &=& \frac{e^{-\lambda}}{8 r} \Big[2 a \nu'-a \lambda ' \left(r \nu'+2\right)+a r \left(2 \nu ''+\nu '^2\right) \nonumber\\
&& -2 b r e^{\lambda} \Big]. \label{eq23}
\end{eqnarray}
where $\rho$ is the energy density with  $p_r$ and $p_t$ are the radial and tangential pressure of the matter sector, considered correspond to a anisotropic fluid.

\begin{figure}[t]
    \centering
        \includegraphics[scale=.8]{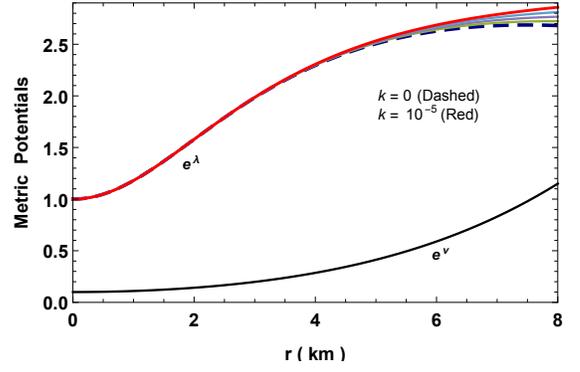}
       \caption{Variation of metric potentials with radial coordinate for $a=1,~b=0.01,~c=0.01,~d=0.0001$ and $A=0.1$ ($f(T)=aT+b$ and T1).}\label{me}
\end{figure}

\begin{figure}[t]
    \centering
        \includegraphics[scale=.8]{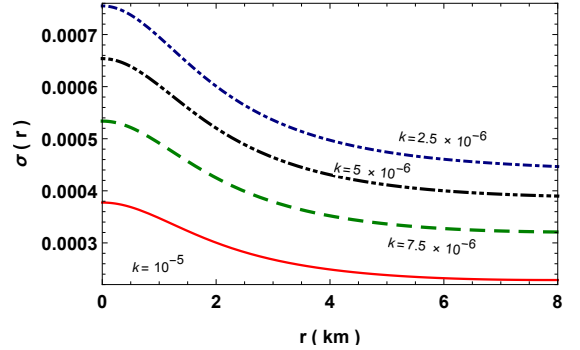}
       \caption{Variation of charge density with radial coordinates ($f(T)=aT+b$ and T1).}\label{sigm}
\end{figure}

\subsection{Off-diagonal tetrad and $f(T)=aT+b$}
 {Proceeding forward the above discussed linear model of $f(T)$ function, we consider another possible tetrad field which is off-diagonal (T2), given by} \cite{Bohmer:2011si}

\begin{eqnarray}
e^i_\mu =\begin{bmatrix}
e^{\nu/2} &      0                                 & 0 & 0 \\
0         &     e^{\lambda/2}\sin \theta \cos \phi & r \cos \theta \cos \phi & -r \sin \theta \sin \phi \\
0         &     e^{\lambda/2}\sin \theta \sin \phi & r \cos \theta \sin \phi & r \sin \theta \cos \phi \\
0         &     e^{\lambda/2} \cos \theta & -r \sin \theta & 0\\
\end{bmatrix}.\nonumber
\end{eqnarray}

\begin{figure}[t]
    \centering
        \includegraphics[scale=.8]{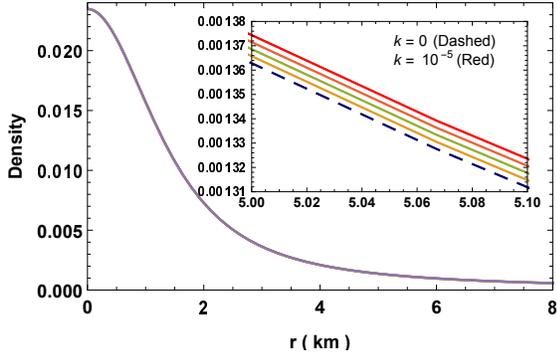}
       \caption{Variation of density with radial coordinate ($f(T)=aT+b$ and T1).}\label{de}
\end{figure}

\begin{figure}[t]
    \centering
        \includegraphics[scale=.8]{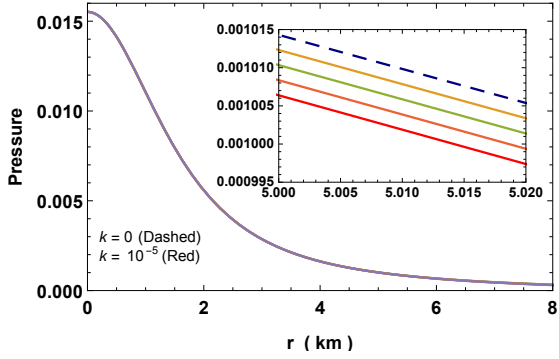}
       \caption{Variation of pressure with radial coordinate ($f(T)=aT+b$ and T1).}\label{pr}
\end{figure}

By doing a rotation, the off-diagonal basis tetrad is related to its diagonal form.  One can obtain $e$ = $|e^i_\mu| = r^2 \sin \theta ~ e^{(\nu+\lambda)/2}$, and we determine the torsion  scalar as
\begin{eqnarray}
T(r) = {2e^{-\lambda} \over r^2}\Big(e^{\lambda/2}-1 \Big) \Big(e^{\lambda/2}-1-2 \nu' \Big).
\end{eqnarray}
Inserting this and the components of the
tensors $S$ and $T$ into the equation (\ref{eq7}), we obtain the modified field equations
\begin{eqnarray}
4\pi \rho&=& \frac{e^{-\lambda}}{4 r^2} \Big[2 a r \lambda '+2 a e^{\lambda /2} \left(r \nu '+2\right)-2 a+b e^{\lambda } r^2 \Big], \nonumber \\
&& - E^2  \label{eq25} \\
4\pi p_r &=& \frac{e^{-\lambda }}{4 r^2} \Big[a \left(2-4 e^{\lambda /2}\right)-2 a \left(e^{\lambda /2}-1\right) r \nu '-b e^{\lambda } r^2 \Big] \nonumber \\
 && +E^2, \label{pr1} \\
4\pi p_t &=& \frac{e^{-\lambda }}{8 r} \Big[2 a \nu '-a \lambda ' \left(r \nu '+2\right)+2 a r \nu '' \nonumber \\
&& +a r \nu '^2-2 b e^{\lambda } r \Big]-E^2. \label{eq27}
\end{eqnarray}

 {We would like to mention that gravitational sector of TEGR is Lorentz invariant in the sense that any choice of the tetrad fields leads to the same equations of motion. Here, we would like to emphasize that the claim made above concerned solely with the argument.}

\subsection{Diagonal tetrad and $f(T)=aT^2$}

Proceeding forward and using the T1 tetrad with  $f(T)=aT^2$, one can get the following field equations:

\begin{figure}[t]
    \centering
        \includegraphics[scale=.8]{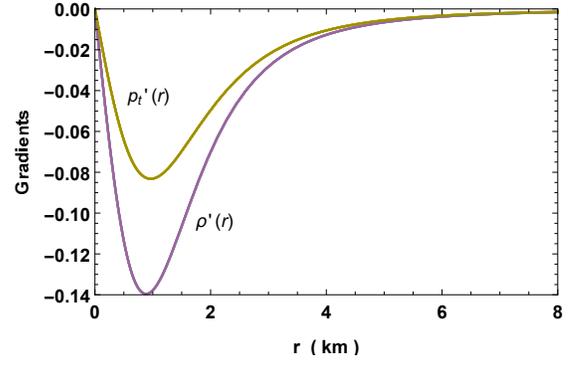}
       \caption{Variation of density and pressure gradients with radial coordinate ($f(T)=aT+b$ and T1).}\label{gr}
\end{figure}

\begin{figure}[t]
    \centering
        \includegraphics[scale=.8]{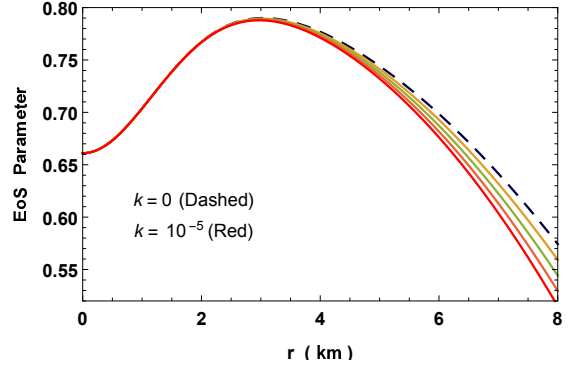}
       \caption{Variation of equation of state (EoS) parameter with radial coordinate ($f(T)=aT+b$ and T1).}\label{eo}
\end{figure}

\begin{eqnarray}
4\pi \rho &=& \frac{a e^{-2 \lambda }}{r^4} \left(r \nu '+1\right) \left(2 e^{\lambda }+2 r \lambda '-r \nu '-3\right)\label{eq28} \nonumber \\
&& -E^2,\\
4\pi p_r &=& \frac{a e^{-2 \lambda}}{r^4} \left(r \nu '+1\right) \left(-2 e^{\lambda }+3 r \nu '+3\right) +E^2, \label{eq29}\\
4\pi p_t &=& \frac{a e^{-2 \lambda }}{2 r^4} \left(r \nu '+1\right) \Big[r \big\{\nu '-\lambda ' (r \nu '+2)+2 r \nu '') \nonumber \\
&& (r \nu '+4)\big\}+2\Big]-E^2.\label{eq30}
\end{eqnarray}
Additional information are required to solve the above field equations.

\subsection{Off-diagonal tetrad and $f(T)=aT^2$}

Taking into account the off-diagonal tetrad and viable power-law form of the $f(T)=aT^2$ model, the field equations reduce to
\begin{eqnarray}
4\pi \rho &=& \frac{a e^{-2 \lambda}}{r^4} \left(1-e^{\lambda/2}\right) \left(e^{\lambda/2}-r \nu '-1\right) \Big[3-2 r \lambda ' \nonumber \\
&& \{r \nu '+2\}+e^{\lambda}+r \nu '-3 e^{\lambda/2}\Big]-E^2,
\end{eqnarray}
\begin{eqnarray}
4\pi p_r &=& \frac{a e^{-2 \lambda }}{r^4} \left(e^{\lambda /2}-1\right) \left(e^{\lambda /2}-r \nu '-1\right) \Big[e^{\lambda }-6 e^{\lambda /2}- \nonumber \\
&& 3 \left\{e^{\lambda /2}-1\right\} r \nu '+3\Big]+E^2, \label{prt2}\\
4\pi p_t &=& \frac{a e^{-2 \lambda (r)}}{2 r^4} \left(e^{\lambda /2}-1\right) \left(e^{\lambda /2}-r \nu '-1\right) \times \nonumber \\
&& \Big[2 \Big\{ \left(e^{\lambda /2}-1\right)^2+r^2 \nu ''\Big\}+r^2 \nu '^2-r \lambda '(r \nu '+2) \nonumber \\
&& -2 \left(e^{\lambda /2}-2\right) r \nu '\Big]-E^2.
\end{eqnarray}
To proceed further, we will assume $p_r=0$ and a specific form of electric field $E$ in the proceeding section.

\subsection{Field equations in pure GR}

The well known field equation in the framework of GR with anisotropic stress-energy tensor profile is given by
\begin{eqnarray}\label{g3}
\rho  &=& \frac{1}{8\pi}\left\{ \frac{1 - e^{-\lambda}}{r^2} + \frac{\lambda'e^{-\lambda}}{r}\right\}, \label{ef} \\
p_r &=& \frac{1}{8\pi} \left\{ \frac{\nu' e^{-\lambda}}{r} - \frac{1 - e^{-\lambda}}{r^2}\right\}, \label{pr3}  \\
 p_t &=& \frac{e^{-\lambda}}{32\pi}\bigg\{2\nu'' + \nu'^{2}  - \nu'\lambda'+ \frac{2\nu'}{r}-\frac{2\lambda'}{r}\bigg\} .                    \label{fe2}
\end{eqnarray}

\begin{figure}[t]
    \centering
        \includegraphics[scale=.8]{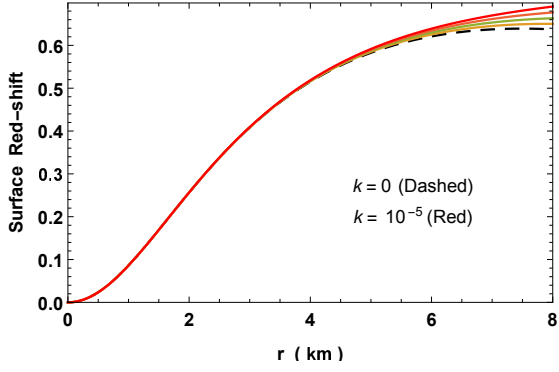}
       \caption{Variation of surface red-shift with radial coordinate ($f(T)=aT+b$ and T1).}\label{re}
\end{figure}

\begin{figure}[t]
    \centering
        \includegraphics[scale=.8]{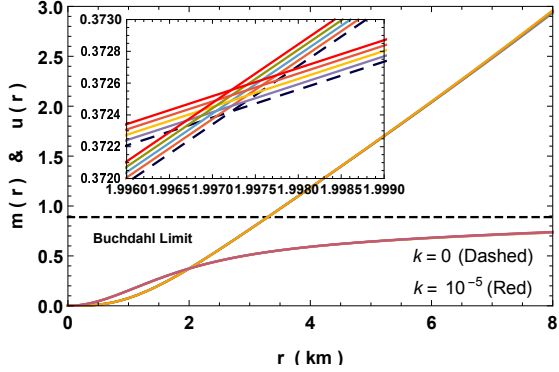}
       \caption{Variation of mass and radius with radial coordinate ($f(T)=aT+b$ and T1).}\label{ma}
\end{figure}

\section{Einstein's cluster in $f(T)-$gravity and pure GR}\label{sec2}

Now, we want to show how it is possible to obtain an Einstein's cluster solution starting from a spherically symmetric metric in the absence and presence of electric charge by considering diagonal/off-diagonal tetrad with Specific $f(T)$ function. More interestingly, we compare these results with the results from pure GR.

\subsection{ Neutral solution with T1 and $f(T)=aT+b$:}

In the case of neutral anisotropic system with vanishing radial pressure \cite{Boehmer:2007az}, Eq. (\ref{eq22}), gives
\begin{equation}
2 a e^{-\lambda} \left(r \nu '+1\right)-2 a-b r^2=0. \label{pro}
\end{equation}

Seeking solutions to the Eq. (\ref{pro}) is extremely difficult due to the presence of two unknown variables. We need at least one additional information. Hence, the simplest conception is to introduce the metric potential of form
\begin{eqnarray}
e^\nu = A+c r^2+d r^4, \label{enu}
\end{eqnarray}
and solving the Eq. (\ref{pro}) using (\ref{enu}), we get
\begin{equation}
e^\lambda = \frac{2 a \left(A+3 c r^2+5 d r^4\right)}{\left(2 a+b r^2\right) \left(A+c r^2+d r^4\right)}.
\end{equation}

Accordingly we obtained solution, the energy density and transverse
pressure in the following form
\begin{eqnarray}
\rho (r) &=& \frac{\left(A+c r^2+d r^4\right) \left(6 a c+20 a d r^2-A b+5 b d r^4\right)}{8\pi \left(A+3 c r^2+5 d r^4\right)^2}, \nonumber \\
\label{rho}\\
p_t(r) &=& \frac{1}{16\pi \left(A+c r^2+d r^4\right) \left(A+3 c r^2+5 d r^4\right)^2} \times \nonumber \\
&& \Big[A^2 \Big\{8 a \left(c+4 d r^2\right)-b r^2 \left(9 c+4 d r^2\right)\Big\}+ \nonumber\\
&& A \Big\{2 a \left(11 c^2 r^2+64 c d r^4+68 d^2 r^6\right)+b r^4 \nonumber\\
&& \left(10 d^2 r^4 -15 c^2-14 c d r^2\right)\Big\}+r^4 \Big\{a \nonumber\\
&& \left(6 c^3+54 c^2 d r^2+96 c d^2 r^4+40 d^3 r^6\right)-b r^2 \nonumber \\
&& \left(12 c^3+37 c^2 d r^2+41 c d^2 r^4+20 d^3 r^6\right)\Big\}- \nonumber \\
&& 2 A^3 b \Big].
\label{pt}
\end{eqnarray}

To examine the system more closely we find the density and pressure gradients takes the form
\begin{eqnarray}
{d\rho \over dr} &=& -\frac{r \big[A f_1(r)-f_2(r)+f_3(r)\big]}{\big(A+3 c r^2+5 d r^4\big)^3}, \\
{dp_t \over dr} &=& -\frac{r}{2 \left(A+c r^2+d r^4\right)^2 \left(A+3 c r^2+5 d r^4\right)^3} \nonumber \\
&& \Big[A^3 f_5(r)-A^4 \left(32 a d+5 b c+36 b d r^2\right) + A^2\nonumber \\
&&  r^2 f_4(r)+A r^4 \left\{f_6(r)+f_7(r)\right\}+r^6 f_8(r) \Big],
\end{eqnarray}
where
\begin{eqnarray}
f_1(r) &=& 2 a \big(15 c^2+64 c d r^2+120 d^2 r^4 \big)+b r^2 \big(20 d^2 r^4- \nonumber \\
&& 3 c^2-30 c d r^2 \big),\nonumber \\
f_2(r) &=& A^2 \left(20 a d+5 b c+28 b d r^2\right)+5 b c d r^6 \left(3 c+d r^2\right), \nonumber \\
f_3(r) &=& 2 a \left(9 c^3 r^2+45 c^2 d r^4+100 c d^2 r^6+50 d^3 r^8\right), \nonumber \\
f_4(r) &=& 2 a \big(63 c^3+275 c^2 d r^2+692 c d^2 r^4+572 d^3 r^6 \big)+ \nonumber \\
&& b r^2 \big(20 d^3 r^6-3 c^3-205 c^2 d r^2-250 c d^2 r^4 \big) \nonumber\\
f_5 (r) &=& 2 a \big(17 c^2+8 c d r^2+52 d^2 r^4 \big)-b r^2 \big(15 c^2+ \nonumber \\
&& 172 c d r^2+140 d^2 r^4 \big),  \nonumber
\end{eqnarray}
\begin{eqnarray}
f_6(r) &=& b r^2 \big(15 c^4-30 c^3 d r^2+95 c^2 d^2 r^4+508 c d^3 r^6 \nonumber \\
&& +380 d^4 r^8 \big), \nonumber \\
f_7(r) &=& 2 a \big(63 c^4+406 c^3 d r^2+1295 c^2 d^2 r^4+1808 c d^3 r^6 \nonumber \\
&& +860 d^4 r^8 \big), \nonumber \\
f_8(r) &=& 2 a \big(9 c^5+63 c^4 d r^2+267 c^3 d^2 r^4+525 c^2 d^3 r^6 \nonumber \\
&& +420 c d^4 r^8+100 d^5 r^{10} \big)-b c d r^4 \big(45 c^3+ \nonumber \\
&& 119 c^2 d r^2+67 c d^2 r^4-15 d^3 r^6 \big). \nonumber
\end{eqnarray}

Now, the EoS parameter and surface red-shift can be found as
\begin{eqnarray}
\omega = {p_t \over \rho} \le 1,~~~~z_s = e^{\lambda_s}-1.
\end{eqnarray}

To conclude this section, we report the gravitational mass and compactness parameter by a spherically symmetric source with radius $r$, we get
\begin{eqnarray}
m(r) &=& 4\pi \int_0^r \rho(\zeta)~ \zeta^2 ~d\zeta, \nonumber \\
&=& \frac{r^3 \big[6 a \left(c+2 d r^2\right)-A b+b d r^4\big]}{6 \left(A+3 c r^2+5 d r^4\right)}, \\
u(r) &=& {2m(r) \over r} = \frac{r^2 \big[6 a \left(c+2 d r^2\right)-A b+b d r^4\big]}{3 \left(A+3 c r^2+5 d r^4\right)}.
\end{eqnarray}

\subsection{Charged solution with T1 and $f(T)=aT+b$:}

Using Eq. (\ref{eq22}), we have
\begin{eqnarray}
\frac{2 a e^{-\lambda} \left(r \nu'+1\right)-2 a-b r^2}{4 r^2}+E^2 = 0.
\end{eqnarray}
Follow the assumption (\ref{enu}) along with $E^2=kr^2$ and the solution is easily found as
\begin{eqnarray}
e^\lambda &=& \frac{2 a \left(A+3 c r^2+5 d r^4\right)}{\left(2 a+b r^2-4 k r^4\right) \left(A+c r^2+d r^4\right)},
\end{eqnarray}
and the proper charge density is given by
\begin{eqnarray}
\sigma(r) &=& \frac{3 \sqrt{k}}{4 \sqrt{2} \pi } \sqrt{\frac{\left(2 a+b r^2-4 k r^4\right) \left(A+c r^2+d r^4\right)}{a \left(A+3 c r^2+5 d r^4\right)}}.
\end{eqnarray}
Hence, at $k=0$ the charge solution reduces to the neutral solution. The variations of metric functions and charge density are shown in Figs. \ref{me} and \ref{sigm}. The next step is to determine the energy density and pressure which are given by
\begin{eqnarray}
8\pi \rho &=& \frac{A+c r^2+d r^4}{2 \left(A+3 c r^2+5 d r^4\right)^2} \Big[6 a c+20 a d r^2-A b \nonumber \\
&& +8 A k r^2+5 b d r^4+12 c k r^4 \Big], \\
8\pi p_t &=& \frac{2 A^3 \left(2 k r^2-b\right)+A^2 g_1(r)+A r^2 g_2(r)-r^4 g_3(r)}{4 \left(A+c r^2+d r^4\right) \left(A+3 c r^2+5 d r^4\right)^2}. \nonumber \\
\end{eqnarray}
where
\begin{eqnarray}
g_1(r) &=& 8 a \left(c+4 d r^2\right)-9 b c r^2-4 b d r^4+4 c k r^4-36 d k r^6, \nonumber \\
g_2(r) &=& 2 a \left(11 c^2+64 c d r^2+68 d^2 r^4\right)-r^2 \big[b \big(15 c^2+14 c d  \nonumber \\
&& r^2-10 d^2 r^4\big)+4 k r^2 \left(4 c^2+46 c d r^2+57 d^2 r^4\right)\big], \nonumber \\
g_3(r) &=& r^2 \big[b \big(12 c^3+37 c^2 d r^2+41 c d^2 r^4+20 d^3 r^6\big) \nonumber \\
&& +4 d k r^4 \big(16 c^2+35 c d r^2+15 d^2 r^4\big)\big]-2 a \big(3 c^3+ \nonumber \\
&& 27 c^2 d r^2+48 c d^2 r^4+20 d^3 r^6\big).\nonumber
\end{eqnarray}
The trends of density and pressure are shown in Figs. \ref{de} and \ref{pr}.

The mass function, compactness parameter and surface red-shift can be calculated as
\begin{eqnarray}
m(r) &=& {1 \over 2}\int_0^r \big(8\pi \rho x^2+E^2 x^2 \big) dx+{q^2 \over 2r} \nonumber \\
&=& \frac{r^3 }{12 \left(A+3 c r^2+5 d r^4\right)} \Big[6 a \left(c+2 d r^2\right) -A  \times \nonumber \\
&& \left(b-12 k r^2\right)+r^4 \left(b d+24 c k+36 d k r^2\right)\Big], \\
u(r) &=& {2m(r) \over r}\\
z_s &=& e^{\lambda_R}-1.
\end{eqnarray}
The variations in gradients, equation of state parameter, surface red-shift, mass function and compactness parameter are shown in Figs. \ref{gr}, \ref{eo}, \ref{re} and \ref{ma}.

\subsection{Charged solution with T2 and $f(T)=aT+b$:}

Proceeding the same as in previous section with vanishing radial pressure and assuming $E^2=k r^2$ for Eq. (\ref{pr1}), our model provides that
\begin{eqnarray}\label{eq52}
a(2-4 e^{\frac{\lambda}{2}})-2 a \nu' r (e^{\frac{\lambda}{2}}-1)-b r^2 e^{\lambda}+kr^2 = 0.
\end{eqnarray}

\begin{figure}[t]
\centering
\includegraphics[scale=.65]{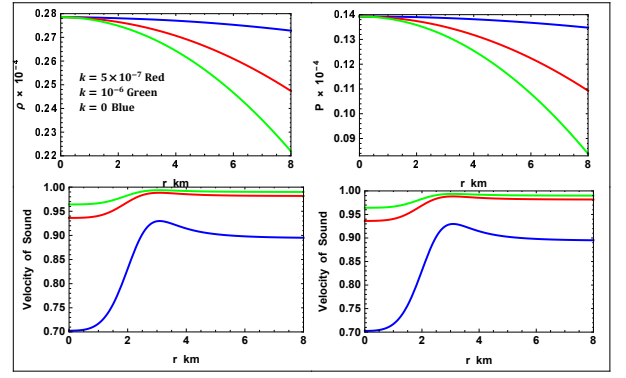}
\caption{Variation of pressure with radial coordinate with for solution in III.3 for $a = 1.5,~b = 0.0007,~c = 0.002,~d = 0.001$ and $A = 0.3$.}\label{td1}
\end{figure}

However, in order to solve Eq. (\ref{eq52}) we assume a specific from of $g_{tt}$ as $e^\nu = A+cr^2+dr^4$, and  the solution can be written as
\begin{eqnarray}
e^\lambda &=& \frac{k}{b}+\frac{2 a \left(A+3 c r^2+5 d r^4\right)}{b r^2 \left(A+c r^2+d r^4\right)}+ \nonumber \\
&& \frac{4 h_3(r)}{b^2 r^4 \left(A+c r^2+d r^4\right)^2} .
\end{eqnarray}
Taking into account the metric potential and plugging those values into the Eqs. (\ref{eq25}-\ref{eq27}), one can easily obtain the stress-energy tensor profile.  However, we avoid to enlist the physical character because of very complicated and lengthy expressions. Alternatively, we could get rid of this complicated expressions through a graphical representation. The qualitative behaviour of the stress-energy tensor components (density, pressure, velocity of sound and adiabatic index) are depicted in Figs. \ref{td1}. { Notice, that the pressure and density are decreasing outward, the velocity of sound is also within the causal limit and the adiabatic index is $> 4/3$. Our approach here follows make sense as a cluster solution which is sufficient to mimic as a compact star. Similar to solution in Sect III.2, one can also see that as charge parameter $k$ increases, the stiffness of the EoS also increases. Therefore, this cluster solution can also mimic properties of compact star}.

\begin{figure}[t]
    \centering
        \includegraphics[scale=.8]{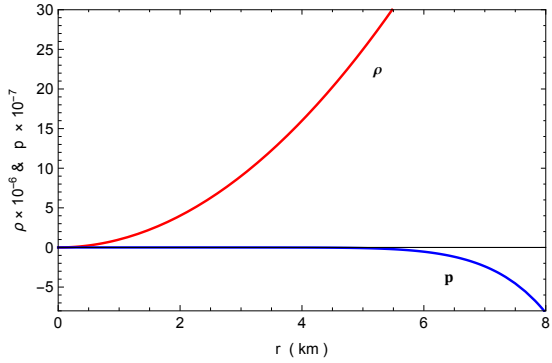}
       \caption{Variation of density and pressure with radial coordinate for in $f(T)=aT^2$ and T1.}\label{td2c}
\end{figure}

\subsection{Charged solution with T1 and $f(T)=aT^2$:}

Further, we introduce the T1 tetrad  and the function $f(T)=aT^2$. For vanishing radial pressure (\ref{eq29}), we get
\begin{equation}
\frac{a e^{-2 \lambda}}{r^4} \left(r \nu '+1\right) \left(3-2 e^{\lambda }+3 r \nu '\right)+E^2=0. \label{ee}
\end{equation}
To solve the Eq. (\ref{ee}), we need an additional information because of the three unknowns $\lambda,~\nu$ and $E$. Therefore, we have assumed the same $e^\nu=A+cr^2+dr^4$ and $E^2=kr^2$. The resulting solution can be written as
\begin{eqnarray}
e^\lambda &=& \frac{1}{2 k} \sqrt{\frac{a^2 [\chi _1(r)]^2}{r^4}-4 a k \chi _2(r)}+\frac{a \chi _1(r)}{r^2},
\end{eqnarray}
where,
\begin{eqnarray}
\chi_1(r) &=& \frac{4 c}{r^2 \left(A+c r^2+d r^4\right)}+\frac{8 d}{A+c r^2+d r^4}+\frac{2}{r^4},\nonumber \\
\chi_2(r) &=& \frac{12 c^2}{r^2 \left(A+c r^2+d r^4\right)^2}+\frac{48 c d}{\left(A+c r^2+d r^4\right)^2}+\frac{3}{r^6} \nonumber \\
&& + \frac{48 d^2 r^2}{\left(A+c r^2+d r^4\right)^2}+\frac{24 d}{r^2 \left(A+c r^2+d r^4\right)} \nonumber \\
&& +\frac{12 c}{r^4 \left(A+c r^2+d r^4\right)}. \nonumber
\end{eqnarray}
Again, we avoid to write the expressions for pressure and density because of their lengthy expressions. Interestingly enough, from the Fig. \ref{td2c}, that the transverse pressure and density vanishes at the center. In spite of the fact that there is no physical solution exist because of the density vanishes at the center and increasing outward.

A crucial point in this discussion is about the neutral solution i.e. $E^2=0$. According to Eq. (\ref{ee}) and  $E^2=0$, the expression is a product of two terms. Investigating solutions for metric potential with the first equality of (\ref{ee}) we obtain $\nu(r)=B-\ln r$. Substituting this value we get $e^\lambda=0$. Therefore, one can immediately conclude that no physical solutions exist in this scenarios.

Furthermore, for T2 tetrad with the same function we found the exactly same situation for neutral case i.e. $\nu=A-\ln r$ and $e^\lambda=0$. Therefore, the main drawback of
the $f(T)=aT^2$ gravity model along the T1 and T2 tetrad is that the obtained solutions do not process any physically realistic Einstein's cluster solution.

\begin{figure}[t]
    \centering
        \includegraphics[scale=.8]{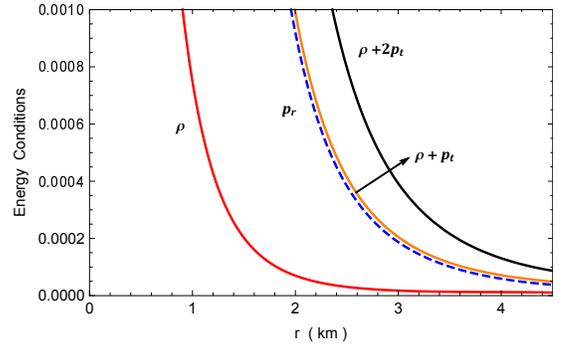}
       \caption{Variation of density, pressure and energy conditions with radial coordinate in T2 and $f(T)=aT^2$.}\label{t2r}
\end{figure}

\begin{figure}[t]
    \centering
        \includegraphics[scale=.8]{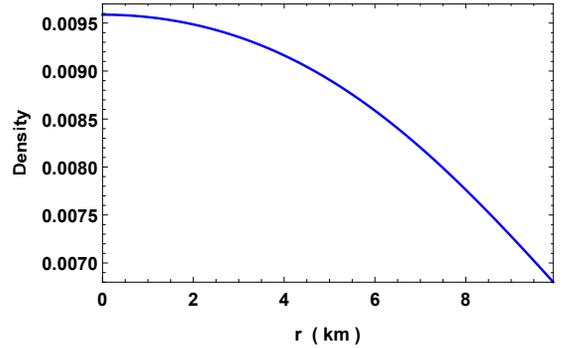}
       \caption{Variation of density with radial coordinate in pure GR.}\label{gt1}
\end{figure}

\begin{figure}[t]
    \centering
        \includegraphics[scale=.8]{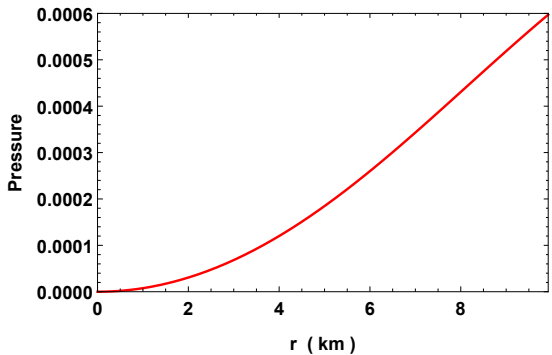}
       \caption{Variation of pressure with radial coordinate in pure GR.}\label{gt2}
\end{figure}

\subsection{Charged solution with T2 and $f(T)=aT^2$:}

For Einstein's cluster, the vanishing radial pressure gives
\begin{eqnarray}
&& \frac{a e^{-2 \lambda }}{r^4} \left(e^{\lambda /2}-1\right) \left(e^{\lambda /2}-r \nu '-1\right) \Big(e^{\lambda }-6 e^{\lambda /2}- \nonumber \\
&& 3 \left\{e^{\lambda /2}-1\right\} r \nu '+3\Big)+E^2=0.
\end{eqnarray}
Here, to simplify the solution we ansatz $e^\nu = A+c r^2$ and $E^2=kr^2$, and the solution gives
\begin{eqnarray}
e^{\lambda/2} &=& \frac{2 a \tau (r)}{e^{\nu } \gamma (r)}+\zeta (r) \bigg[\frac{1}{e^{2 \nu } \gamma (r)^2} \bigg\{  8 a^2 \tau (r)^2-a \gamma (r) \big(4 A^2 \nonumber \\
&& +16 A c r^2+15 c^2 r^4\big)-a e^{\nu } \gamma (r) (A+3 c r^2) + \nonumber \\
&& \frac{e^{\nu } \gamma (r) \zeta (r)}{\tau (r)} \Big[a \left(3 A^2+12 A c r^2+13 c^2 r^4\right)-k r^6 \nonumber \\
&& \left(A^2+4 A c r^2+3 c^2 r^4\right) \Big]\bigg\} \bigg]^{1/2},
\end{eqnarray}
where,
\begin{eqnarray}
\tau(r) &=& A+2 c r^2 ~~,~~ \gamma(r) = a+k r^6 \nonumber\\
\zeta(r) &=& \sqrt{\frac{a \left(a-3 k r^6\right) \left(A+2 c r^2\right)^2}{\left(a+k r^6\right)^2 \left(A+c r^2\right)^2}} ~~. \nonumber
\end{eqnarray}

Due to extremely lengthy expressions of density and pressure will exclude their expressions, however, their
properties have been obtained by numerically integration
for the charged fluid equation of state and by graphical representation. As one can see from Fig. \ref{t2r}, that the pressure and density are positive and decreasing outward, however, they blows up at $r=0$.  Such a solution can mimic compact stars which contain a central singularity. Even if such solutions could exist, they do not gravitationally stable.

\subsection{Neutral Einstein's cluster solution in pure GR}

Up to now, we have concentrated our discussion on modified teleparallel gravity or $f(T)$ gravity. We attempt to discuss here the general relativity case. Let us now concentrate on Eq. (\ref{pr3}), with vanishing radial pressure, we find
\begin{eqnarray}
&& {\nu' e^{-\lambda} \over r}-{1-e^{-\lambda} \over r^2}=0 \nonumber \\
\mbox{or} ~~~~&& e^\lambda = 1+r\nu'. \label{gg}
\end{eqnarray}
Our focus is to obtain a complete solution, and for that we use $e^\nu=A+cr^2+dr^4$ as previously discussed. Then, we find
\begin{eqnarray}
e^\lambda = \frac{A+3 c r^2+5 d r^4}{A+c r^2+d r^4}.
\end{eqnarray}

From Eqs. (\ref{g3}-\ref{fe2}), we deduce
\begin{eqnarray}
8\pi \rho &=& \frac{2 \left(3 c+10 d r^2\right) \left(A+c r^2+d r^4\right)}{\left(A+3 c r^2+5 d r^4\right)^2},\\ \label{eq58}
8\pi p_t &=& \frac{3 c^2 r^2+16 c d r^4+20 d^2 r^6}{\left(A+3 c r^2+5 d r^4\right)^2}. \label{eq59}
\end{eqnarray}

Since, our goal now is to build a more realistic model of Einstein's cluster. To do so we first fix the values of few unknown parameters by matching the  Schwarzschild's vacuum solution at the boundary, which are found to be
\begin{eqnarray}
A &=& \frac{c R^2 (R-3 M)+d R^4 (2 R-5 M)}{M},\\
d &=& \frac{M-c R^3}{2 R^5}.
\end{eqnarray}
The parameter $c$ will be treated as free parameter for tuning purpose. Moreover, $c$ is directly related with determining cluster
solution.

Considering Eq. (\ref{eq59}), one immediately finds that $p_t =0$ at the center $r=0$,
which violates the physical condition of a compact star.
The qualitative behaviour of the density and pressure are depicted in \ref{gt1} and \ref{gt2}.
Note that the energy density is positive throughout the spacetime, but the pressure is zero at the center and increasing outward. In a recent paper,  Thirukkanesh \textit{et al} \cite{thiru} investigated a particular class of stellar solutions  which describe spherically symmetric matter distributions with vanishing radial stresses within the framework of GR.  However, their model process $p_t >0$  everywhere inside the star but it decreases monotonically from the center and reaches a minimum at certain radius, and thereafter increasing monotonically toward the boundary of the star. But our pressure is strictly increasing throughout the interior spacetime. Qualitatively, we verify that for increasing nature of the pressure give raise to imaginary sound speed and the negative values of adiabatic index. In general, this scenario does not process a physically valid compact star. Thus, in pure GR the solutions representing Einstein cluster can't mimic compact stars.

\section{Matching of boundary for charged solution  for diagonal tetrad in linear $f(T)$}\label{sec3}

Having derived the equations that describe charged Einstein cluster, we now proceed to match the interior solution with an exterior Reissner-Nordstr$\dot{\text{o}}$m vacuum solution. Moreover, we fix the values of constant $a,~b,~ A$ and $B$ from junction conditions imposed on the internal and external metrics at the hyper-surface. The Reissner-Nordstr$\dot{\text{o}}$m metric is given by
\begin{eqnarray}
ds^{2}&=&\left(1-\frac{2m}{r}+{q^2 \over r^2}\right)dt^{2}-\left(1-\frac{2m}{r}+{q^2 \over r^2}\right)^{-1}dr^{2}\nonumber \\
&& -r^{2}(d\theta^{2}+\sin^{2}\theta d\phi^{2}) .
\end{eqnarray}
Now, using the continuity of the metric coefficients $e^{\nu}$ and $e^{\lambda}$ across the boundary $r=R$, we get the following
\begin{eqnarray}
\left(1-\frac{2M}{R}+{Q^2 \over R^2}\right)^{-1} &=& \frac{2 a \left(A+3 c R^2+5 d R^4\right)}{\left(2 a+b R^2-4 k R^4\right) } \nonumber \\
&& \times \left(A+c R^2+d R^4\right)^{-1}, \label{b1}\\
1-\frac{2M}{R}+{Q^2 \over R^2} &=& A+c R^2+d R^4. \label{b2}
\end{eqnarray}

Solving Eqs. (\ref{b1}) and (\ref{b2}), we get
\begin{eqnarray}
a &=&  R^2 \left(b-4 k R^2\right) \left(A+c R^2+d R^4\right) \Big[2 A k R^4-2 A U+ \nonumber \\
&& 6 c k R^6-6 c R^2 U+4 c R^2+10 d k R^8-10 d R^4 U \nonumber \\
&& +8 d R^4 \Big]^{-1},\\
A &=& 1 - c R^2 - d R^4 + k R^4 - {2M \over R}.
\end{eqnarray}
Here $U=2M/R$, $c,~d, ~k$ and $b$ will be treated as fitting parameters while $M$ and $R$ can be chosen from observed values of compact stars.

\section{Non-singular nature of the solution (Sect III.2)}\label{sec4}

The central density and central pressure are surprisingly independent of electric charge and found as
\begin{eqnarray}
8\pi \rho_c &=& \frac{6 a c-A b}{2 A} > 0, \label{roc}\\
8\pi p_{tc} &=& \frac{4 a c- A b}{2 A} > 0, \label{pc}
\end{eqnarray}
which implies $4ac > Ab$. The solution also satisfies the Zeldovich's \cite{zeld} condition as
\begin{equation}
{p_{tc} \over \rho_c} = {4ac-Ab \over 6ac -Ab} < 1.
\end{equation}
Therefore, the solution doesn't contain any singularity and also can represent physical matters.

\begin{figure}[t]
    \centering
        \includegraphics[scale=.6]{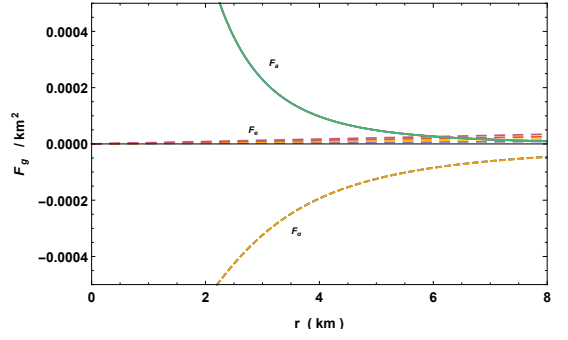}
       \caption{Variation of forces acting on TOV-equation with radial coordinate ($f(T)=aT+b$ and T1).}\label{to}
\end{figure}

\section{Equilibrium and stability analysis (Sect. III.2)}\label{sec5}

The task is now to study the stability of self-consistent regular solution. In the present section, we analyse the stability of the Einstein clusters by performing some analytical calculations and plotting several figures.

\subsection{Equilibrium analysis via TOV-equation}
In the spirit of completion we discuss the stability of the Einstein cluster model. Consider hydrostatic equilibrium via Tolman-Oppenheimer-Volkoff (TOV) equation. Now by employing the generalized-TOV equation \cite{Tolman,Oppenheimer} in the presence of charge, as prescribed in \cite{Ponce}, we have the following form
\begin{equation}
-\frac{M_g(r)~\rho(r)}{r}e^{(\nu-\lambda)/2}+\frac{2p_t(r)}{r}+\sigma E e^{\lambda/2}=0, \label{tov1}
\end{equation}
where $M_g(r) $ represents the gravitational mass within the radius $r$. It is defined using the Tolman-Whittaker mass formula through the Einstein's field equations as

\begin{eqnarray}
M_g(r) &=& 4 \pi \int_0^r \big(T^t_t-T^r_r-T^\theta_\theta-T^\phi_\phi \big) r^2 e^{(\nu+\lambda)/2}dr, \nonumber\\
&=& \frac{1}{2}re^{(\lambda-\nu)/2}~\nu'. \label{mg}
\end{eqnarray}

Now, plugging the value of $M_g(r)$ in equation (\ref{tov1}), we get
\begin{eqnarray}
&& -\frac{\nu'}{2}~\rho(r)+\frac{2p_t(r)}{r}+\sigma E e^{\lambda/2}=0 \nonumber\\
\mbox{or}~ &&  F_g+F_a+F_e = 0,
\end{eqnarray}
where $F_g, ~F_e$ and $F_a$ are the three different forces namely gravitational, anisotropic and electromagnetic forces, respectively. For our system the forces are as follows:
\begin{eqnarray}
F_g = -\frac{\nu'}{2}~\rho(r),~~~~
F_a = {2p_t(r) \over r} ,~~~~
F_e = \sigma E e^{\lambda/2}.
\end{eqnarray}
The variation of forces in TOV-equation w.r.t. the radial coordinate $r$ is given in Fig. \ref{to}.

\subsection{Causality condition and stability criterion}

For static spherically symmetric spacetime solution one has to check also the behavior of speed of sound propagation $v^2$, which is given by the expression $dp/d\rho$. Normally it is believed that the velocity of sound is less than the velocity of light. For that the speed of sound should be $\le 1$ and it can be determined as

\begin{figure}[t]
    \centering
        \includegraphics[scale=.8]{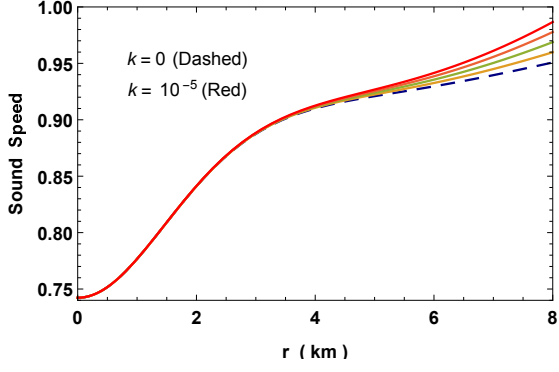}
       \caption{Variation of sound speed with radial coordinate ($f(T)=aT+b$ and T1).}\label{so}
\end{figure}

\begin{eqnarray}
v_t^2 &=& {dp_t \over d\rho }= \frac{1}{2 \chi _6(r) \left(A+c r^2+d r^4\right)^2} \times \nonumber \\
&& \Big[4 A^5 k+A^4 \chi _1(r)-A^3 \chi _2(r)-A^2 r^2 \chi _3(r)- \nonumber \\
&& A r^4 \chi _4(r)+r^6 \chi _5(r) \Big],
\end{eqnarray}
where,
\begin{eqnarray}
\chi _1(r) &=& 32 a d+b \left(5 c+36 d r^2\right)-4 k r^2 \left(c+43 d r^2\right), \nonumber \\
\chi _2(r) &=&  2 a (17 c^2+8 c d r^2+52 d^2 r^4)-b (15 c^2 r^2+172 c d \nonumber \\
&& r^4+140 d^2 r^6)+4 k r^4 (17 c^2+268 c d r^2+274 d^2 r^4), \nonumber \\
\chi _3(r) &=& 2 a (63 c^3+275 c^2 d r^2+692 c d^2 r^4+572 d^3 r^6)- \nonumber \\
&& b (3 c^3 r^2+205 c^2 d r^4+250 c d^2 r^6-20 d^3 r^8)+4 k r^4 \nonumber \\
&& (23 c^3+499 c^2 d r^2+946 c d^2 r^4+426 d^3 r^6), \nonumber\\
\chi _4(r) &=& 2 a (63 c^4+406 c^3 d r^2+1295 c^2 d^2 r^4+1808 c d^3 r^6+ \nonumber \\
&& 860 d^4 r^8)+b r^2 (15 c^4-30 c^3 d r^2+95 c^2 d^2 r^4+ \nonumber \\
&& 508 c d^3 r^6+380 d^4 r^8)+4 d k r^6 (294 c^3+775 c^2 d r^2 \nonumber \\
&& +516 c d^2 r^4+15 d^3 r^6), \nonumber
\end{eqnarray}
\begin{eqnarray}
\chi _5(r) &=& -2 a (9 c^5+63 c^4 d r^2+267 c^3 d^2 r^4+525 c^2 d^3 r^6+\nonumber \\
&& 420 c d^4 r^8+100 d^5 r^{10})+b c d r^4 (45 c^3+119 c^2 d r^2 \nonumber \\
&& +67 c d^2 r^4-15 d^3 r^6)-4 d k r^6 (96 c^4+363 c^3 d r^2+ \nonumber \\
&& 485 c^2 d^2 r^4+285 c d^3 r^6+75 d^4 r^8), \nonumber \\
\chi _6(r) &=& A^2 (20 a d+5 b c+28 b d r^2+16 c k r^2-96 d k r^4)-A \nonumber \\
&&  \big[2 a(15 c^2+64 c d r^2+120 d^2 r^4)+b (-3 c^2 r^2-30 c \nonumber \\
&&  d r^4+20 d^2 r^6)+4 k r^4 (-9 c^2+32 c d r^2+10 d^2 r^4)\big] \nonumber \\
&& -2 a (9 c^3 r^2+45 c^2 d r^4+100 c d^2 r^6+50 d^3 r^8)+ \nonumber \\
&& 8 A^3 k+c r^6 (3 c+d r^2) (5 b d+12 c k). \nonumber
\end{eqnarray}
The presented solution also satisfy the causality condition (see Fig. \ref{so}). Since there is no radial sound speed, the Herrera's cracking method of analyzing stability is not applicable in Einstein's clusters.

\begin{figure}[t]
    \centering
        \includegraphics[scale=.8]{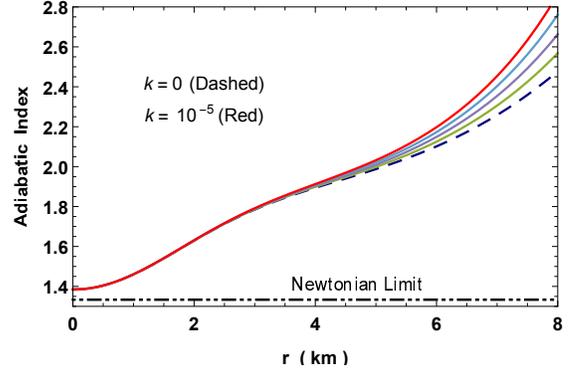}
       \caption{Variation of adiabatic index with radial coordinate ($f(T)=aT+b$ and T1).}\label{ga}
\end{figure}

\begin{figure}[t]
    \centering
        \includegraphics[scale=.8]{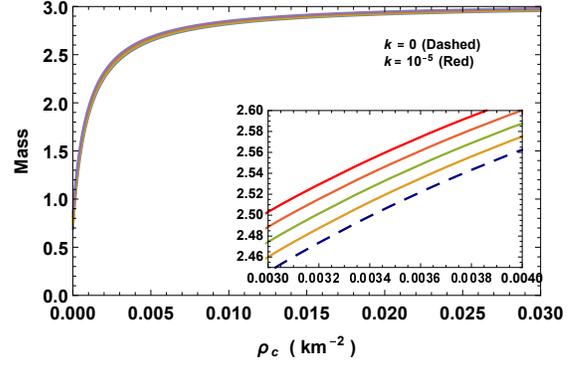}
       \caption{Variation of mass with central density ($f(T)=aT+b$ and T1).}\label{mc}
\end{figure}

\subsection{Stability analysis using relativistic adiabatic index}

Our particular interest is to study the stable equilibrium configuration of spherically symmetric Einstein cluster, and the adiabatic index is a basic ingredient of the stable/instable criterion. Now considering an adiabatic perturbation, the  adiabatic index $\Gamma$, is by \cite{Chandrasekhar,Merafina,chan}
\begin{equation}
\Gamma = {\rho+p_t \over p_t} ~{dp_t \over d\rho},
\end{equation}
where $dp/d\rho$ is the speed of sound in units of speed of light. This approximation leads to a very useful information for compact astrophysical objects and impose some marginal constraints.  In view of the above consideration, Bondi \cite{bond} had clearly mentioned that a stable Newtonian sphere has $\Gamma > 4/3$ and $\Gamma = 4/3$ for a neutral equilibrium. Its values vary from 2 to 4 in most of the neutron stars equations of state \cite{Haensel}. For the solution also the adiabatic index is greater than 4/3. In Fig. \ref{ga} one can see that the central value of the adiabatic index is independent of electric charge and is about 1.386.

\subsection{Static stability criterion}

In order to clarify further the effect of mass-radius and mass-central density relation for the stable stellar configuration, Harrison-Zeldovich-Novikov \cite{harr,zel}
argued that an increasing mass profile with increasing central density i.e. $\partial M/\partial \rho_c > 0$ represents stable configurations and vice-versa. In particular stable or unstable region is achieved when the mass remains constant with increase in central density  i.e. $\partial M/\partial\rho_c = 0$. For the new solution $M(\rho_c)$ and $\partial M/ \partial \rho_c$ is found to be
\begin{eqnarray}
M(\rho_c) &=& \frac{R^3}{30 \left(\frac{6 a c}{b+8 \pi  \rho_c  }+3 c R^2+5 d R^4\right)} \bigg[\frac{6 a c \left(24 k R^2-5 b\right)}{b+8 \pi  \rho_c } \nonumber\\
&& +30 a \left(c+2 d R^2\right)+R^4 (5 b d+12 c k) \bigg],\\
{\partial M \over \partial\rho_c} &=& \frac{24 \pi  a c R^3 \left(c+2 d R^2\right) \left(2 a+b R^2-4 k R^4\right)}{\left(6 a c+R^2 (b+8 \pi  \rho_c  ) \left(3 c+5 d R^2\right)\right)^2}.
\end{eqnarray}
Here, we can clearly see that $\partial M / \partial\rho_c>0$ i.e. mass is increasing function of its central density (Fig. \ref{mc}) and therefore can represents static stable configuration.

\begin{figure}[t]
    \centering
        \includegraphics[scale=.8]{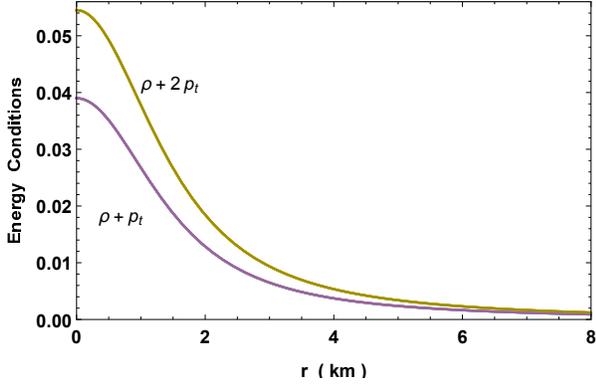}
       \caption{Variation of energy conditions with radial coordinate ($f(T)=aT+b$ and T1).}\label{ec}
\end{figure}

\begin{figure}[t]
    \centering
        \includegraphics[scale=.8]{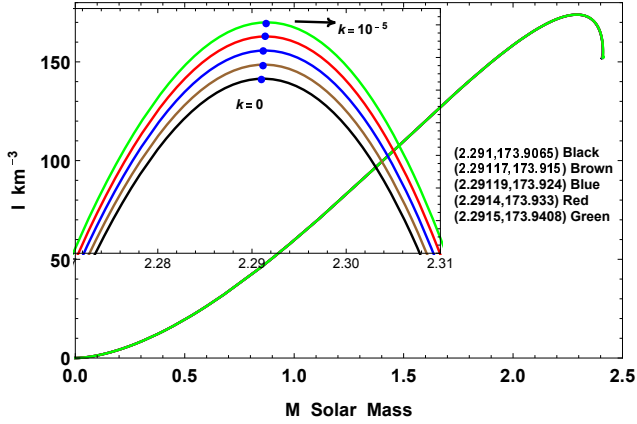}
       \caption{Variation of moment of inertia with mass for $a=1,~b=0.01,~c=0.01,~d=0.0001$ and $A=0.1$ ($f(T)=aT+b$ and T1).}\label{im}
\end{figure}

\begin{figure}[t]
    \centering
        \includegraphics[scale=.8]{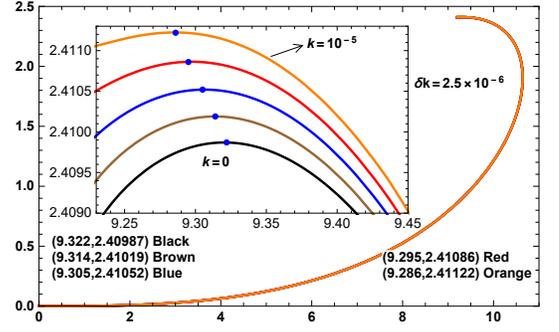}
       \caption{Variation of mass with radius for $a=1,~b=0.01,~c=0.01,~d=0.0001$ and $A=0.1$ ($f(T)=aT+b$ and T1).}\label{mr}
\end{figure}

\begin{figure}[t]
    \centering
        \includegraphics[scale=.7]{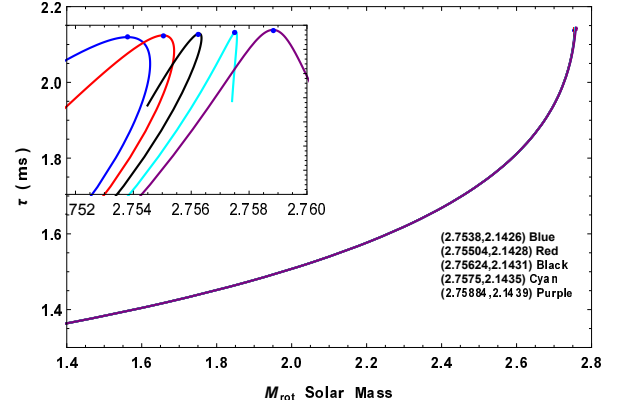}
       \caption{Variation of time period of rotation with mass for $a=1,~b=0.01,~c=0.01,~d=0.0001$ and $A=0.1$ ($f(T)=aT+b$ and T1).}\label{fm}
\end{figure}

\section{Energy conditions}

It is reasonable to expect that models of charged perfect fluids satisfy the energy conditions and such condition depends on the relation between matter density and pressure obeying certain restrictions. In view of the above situation, we examine Strong (SEC), Weak (WEC), Dominant (DEC) and Null (NEC) energy conditions, which are defined as
\begin{eqnarray}
\text{WEC} &:& \rho \geq  0,~\rho+p_t \ge 0, \\
\text{NEC} &:& \rho+p_t \geq  0,\\
\text{DEC} &:& \rho \ge |p_t|,\\
\text{SEC} &:& \rho+2 p_t \ge 0.
\end{eqnarray}
Using the above expression, one can easily justify the nature of energy conditions. The presented model also fulfill these energy conditions (Fig. \ref{ec}). Since the allowed values of the electric charge parameter $k$ is very small i.e. $0\le k \le 1.3 \times 10^{-5}$, the graphs of the above energy conditions are very closed to each other.

\section{Slow rotation model, moment of inertia and time period}\label{sec6}

For a uniformly rotating star with angular velocity $\Omega$, the moment of inertia is given by \cite{litt}
\begin{eqnarray}
I = {8\pi \over 3} \int_0^R r^4 (\rho+p_r) e^{(\lambda-\nu)/2} ~{\omega \over \Omega}~dr,
\end{eqnarray}
where, the rotational drag $\omega$ satisfy the Hartle's equation \cite{hart}
\begin{eqnarray}
{d \over dr} \left(r^4 j ~{d\omega \over dr} \right) =-4r^3\omega~ {dj \over dr} .
\end{eqnarray}
with $j=e^{-(\lambda+\nu)/2}$ which has boundary value $j(R)=1$. The approximate solution of moment of inertia $I$ up to the maximum mass $M_{max}$ was given by Bejger and Haensel \cite{bej} as
\begin{equation}
I = {2 \over 5} \left(1+ { (M/R)\cdot km \over M_\odot}\right) {MR^2},
\end{equation}
The $M-I$ graph is shown in Fig. \ref{im}. The corresponding $M-R$ graph is also shown in Fig. \ref{mr}. From these two graphs we can see that the maximum moment of inertia and maximum mass increases with increase in electric charge. However, the mass corresponding to $I_{max}$ from $M-I$ graph is less by about 5\% as compare to $M_{max}$ from $M-R$ graph. This suggest that the corresponding equation of state is free from softening due to hyperonization or phase transition to an exotic state \cite{buli}.

The minimum time-periods of any rotating compact stars can be expressed with good precision in terms of the masses and radii of the non-rotating configurations. So long as the equation of states obeyed subluminal sound speeds one can expressed the most accurate minimum time period as \cite{ase}
\begin{equation}
\tau \approx 0.82 \left({M_\odot \over M} \right)^{1/2} \left({R \over 10 ~km} \right)^{3/2} ~ms.
\end{equation}
The maximum values of each minimum time periods are almost equal and are negligibly affected by the presence of electric charge (Fig. \ref{fm}). \\

\section{Results and discussions}\label{sec7}

In this paper, we present a model of Einstein's cluster mimicking compact star in the context of TEGR, the Teleparallel Equivalent of General Relativity, as a gauge theory of translations with the torsion tensor being non-zero but with a vanishing curvature tensor, hence, the manifold is globally flat. Considering Einstein's clusters in GR realm arises many un-physical outcomes, such as pressure increases outward, imaginary sound speed, negative adiabatic index and therefore can't mimic compact star model (see discussion in III.6). We have developed the TEGR field equations having a diagonal and off-diagonal tetrad with a specific function of $f(T)$. { More specifically, considering the field equations with a diagonal (T1) and off-diagonal (T2) tetrads with linear functional form of $f(T)=aT+b$, we found Einstein cluster solutions that behaves like a compact star. Thus, it seems interesting that relativistic star solutions are possible only in the case of teleparallel equivalent of general relativity.} In connection with this, we have other solutions for particular power-law of $f(T)$ model with diagonal and off-diagonal tetrad.  However, most of the attempts are unsuccessful because the resulting solutions yield negative pressures. Indeed, we found a very compact cluster solution in the case of $f(T)=aT^2$ using a off-diagonal tetrad. In this case we found the decreasing and positive energy density and pressure, however, both blows up at $r=0$ i.e. it contain a central singularity which is unstable under gravitational collapse. This may be because of the fact that the constraint on the field equations i.e. $f_{,TT}=0$ puts a strict restriction on the choice of $f(T)$ function to a linear one.  As a result, only if $f(T)$  is a linear function of the torsion scalar $T$, one can leads to the existence of neutron star solution \cite{Deliduman:2011ga}.  In a recent paper, \cite{Boehmer:2011gw,Deliduman:2011ga} suggested that, instead of choosing $f_{,TT}=0$, if one  consider $T'=0$ or $T=T_0$, the solution yields a constant energy density and pressure, obeying the dark energy equation of state or the pressure which blows up at $r\rightarrow 0$. This result is similar to our solution for $f(T)=aT^2$ in T2. Therefore, such solutions can't be used to model neutron star alike cluster solution.  \\

For the case of $f(T)=aT+b$ in T1 (see section  III.1 and III.2) two solutions of clusters were found. As per the rigorous analysis and figures, presented model satisfy causality, energy condition, TOV-equation, Bondi criterion and stable static criterion. This means that the solution has the ability to mimic compact star models. The $M-I$ and $M-R$ graphs suggested that the $I_{max}$ and $M_{max}$ increases with increase in charge parameter $k$. The $M-\rho_c$ graph signifies that the solution gain its stability with increase in electric charge. However, the maximum time-period of rotation $\tau_{max}$ is negligibly affected by the presence of electric charge. The stiffness in the equation of state seems to be same and independent of electric charge from center till upto about 3.5 km, however, beyond 3.5 km till the surface, the stiffness increases with increase in electric charge. This may be because of the central region ($0\le r \le 3.5$ km) is extremely dense thus neutralizing the electric charge through $e+p\rightarrow n+\nu_e$, which may also the source of neutrinos as described in \cite{Hogan}. As the density decreases outward, the gravity becomes slightly weaker and the repulsive electric field starts  affecting the stiffness. The solution favor physical solution for the range $0 \le k \le 1.3 \times 10^{-5}$, beyond which the solution doesn't satisfy causality and trigger a gravitational collapse once crossed the Buchdahl limit if $k>1.3 \times 10^{-5}$. { Similarly, for the case of linear $f(T)$ in T2, we have also found cluster solution which mimic the nature of compact star. The solution gives physical cluster solution for a very narrow range of charge parameter $k$ which must be in the range $0< k \le 10^{-6}$ or otherwise the solution violates the causality condition or physically unacceptable.} Overall, the presented solution with vanishing radial pressure and/or Einstein's cluster model is fit for mimicking compact star models.

\subsection*{Acknowledgments}
Farook Rahaman  would like to thank the authorities of the Inter-University Centre for Astronomy and Astrophysics, Pune, India for providing the research facilities.  FR is  also thankful to DST-SERB,  Govt. of India and RUSA 2.0, Jadavpur University,  for financial support. We would also like to thank referee for his or her valuable comments.

\end{document}